\documentclass{aa}
\usepackage{units}
\usepackage{url}
\usepackage{natbib}
\bibpunct{(}{)}{;}{a}{}{,}

\usepackage{color}
\usepackage{hyperref}
\usepackage{graphicx}

\begin{document}
\title{VLBI observations of Jupiter with the Initial Test Station of LOFAR and the Nan\c{c}ay Decametric Array}

\author{A. Nigl\inst{1} \and P. Zarka\inst{2} \and J. Kuijpers\inst{1}
\and H. Falcke\inst{3} \and L. B\"ahren\inst{3} \and L. Denis\inst{2}}

\institute{Department of Astrophysics (IMAPP), Radboud University Nijmegen,
6535~ED~Nijmegen, The~Netherlands \and LESIA, CNRS/Observatoire
de Paris, 92195 Meudon, France \and ASTRON, 7990~AA~Dwingeloo,
The~Netherlands}

\abstract
%CONTEXT
{}
%AIMS
{To demonstrate and test the capability of the next generation of low-frequency radio telescopes to perform high resolution observations across intra-continental baselines. Jupiter's strong burst emission is used to perform broadband full signal cross-correlations on time intervals of up to hundreds of milliseconds.}
%METHODS
{Broadband VLBI observations at about 20 MHz on a baseline of $\sim$ 50000 wavelengths were performed to achieve arcsecond angular resolution. \textsc{lofar}'s Initial Test Station (\textsc{lofar/its}, The Netherlands) and the Nan\c{c}ay Decametric Array (\textsc{nda}, France) digitize the measured electric field with 12 bit and 14 bit in a 40 MHz baseband. The fine structure in Jupiter's signal was used for data synchronization prior to correlation on the time-series data.}
%RESULTS
{Strong emission from Jupiter was detected during snapshots of a few seconds and detailed features down to microsecond time-scales were identified in dynamic spectra. Correlations of Jupiter's burst emission returned strong fringes on 1 ms time-scales over channels as narrow as a hundred kilohertz bandwidth.}
%CONCLUSIONS
{Long baseline interferometry is confirmed at low frequencies, in spite of phase shifts introduced by variations in ionospheric propagation characteristics. Phase coherence was preserved over tens to hundreds of milliseconds with a baseline of $\sim$ 700 km. No significant variation with time was found in the correlations and an estimate for the fringe visibility of 1, suggested that the source was not resolved. The upper limit on the source region size of Jupiter Io-B S-bursts corresponds to an angular resolution of $\sim$ 3 arcseconds. Adding remote stations to the \textsc{lofar} network at baselines up to thousand kilometers will provide 10 times higher resolution down to an arcsecond.}

\authorrunning{Nigl et al.}
\titlerunning{Jupiter VLBI with LOFAR/ITS and NDA}
\offprints{A. Nigl}
\mail{anigl@astro.ru.nl}
\date{Received 31 January 2007 / Accepted 31 May 2007}
\maketitle

\section{Introduction\label{sec:jupintro}}

A new wave of low-frequency telescopes are about to enter astronomy.
$\textsc{lofar}$ (low frequency array) is a new-generation giant
digital telescope and already its prototype ($\textsc{lofar/its}$)
is capable of performing broadband and phase-sensitive interferometry
with high angular, time and frequency resolution. In this paper, the
nature of Jupiter's burst emission is studied at low frequencies ($<
40~\unit{MHz}$) on a baseline of $\sim$50000 wavelengths and an angular resolution
of $\sim3$ arcseconds. We want to test over which bandwidth and time-scales
coherence in Jupiter burst signal is preserved or can be reconstructed,
in spite of ionospheric propagation effects causing phase variation,
Faraday rotation, dispersion and refraction. We want to demonstrate
that high resolution observations can be performed by cross correlations
of signals from remote stations being added to the $\textsc{lofar}$
telescope at baselines up to a thousand kilometers.

Long baseline interferometry of Jupiter's decametric emission ($\textsc{dam}$)
has been performed since 1965 on baselines up to more than 7000 km,
setting an upper limit on the source size of 0.05 arcseconds \citep{lynch72}.
Research was mainly done in the sixties and seventies using narrow-band
analog recording techniques and 1 bit baseband digitization (\citealt{slee65}
\citealt{dulk67} \citealt{brown68} \citealt{dulk70} \citealt{stannard70}
\citealt{lynch72} \citealt{lynch76}).

Jupiter's $\textsc{dam}$ is produced below 40 MHz, limited by the
maximum electron gyrofrequency at Jupiter's surface. The lower limit
for ground-based observations is set by the plasma frequency of the
Earth ionosphere between 5 MHz and 10 MHz. The emission mechanism
of Jupiter's $\textsc{dam}$, modulated by Io, is believed to be cyclotron
maser emission \citep{wu79}. The emission can be very intense up
to $10^{5-6}$ Jy is beamed into a thin hollow cone with its axis
parallel to the local magnetic field. The hollow cone has a half-angle
of 60$^{\circ}$ to 90$^{\circ}$ relative to the local magnetic field
direction and has a cone wall thickness of a few degrees \citep{kaiser00}.
This emission can be detected on Earth when the line of sight of the
observer coincides with the wall of the hollow cone. The emission
arrives in noise storms called long bursts (L-bursts) and in narrow
millisecond bursts called short bursts (S-bursts). \citet[1986,][]{phillips87}
found a first indication for source growth during L-bursts from smaller
than 10 arcseconds up to 60 arcseconds. For S-bursts a complex microstructure
at typically $\sim$40 \ensuremath{µ}s period and $\sim$30 kHz bandwidth
has been observed. These sub-burst wave packets have been attributed
to clouds of electrons moving along the Io flux tube ($\textsc{ift}$)
\citep{carr99}.

\citet{dulk70} put an upper limit of 400 km to 4000 km at Jupiter
on the size of the $\textsc{dam}$ source at 34 MHz, which corresponds
to an angular resolution of $0.1^{"}$ to $1^{"}$. S-burst analysis
provides an upper limit of about 20 km on the instantaneously emitting
source \citep{zarka96}, which translates to 6 milli-arcseconds at
the distance of 4.25 AU.

Jupiter's $\textsc{dam}$ emission is strongly elliptically polarized.
The signal at frequencies near 20 MHz is on average 49\% circularly
and 87\% linearly polarized (\citealt{shaposhnikov97} \citealt{queinnec01}).

For the following analysis, Jupiter was observed with the two radio
telescopes $\textsc{lofar/its}$ (Initial Test Station of $\textsc{lofar}$,
The Netherlands) and $\textsc{nda}$ (Nan\c{c}ay Decametric Array,
France), which are described in Sect. \ref{sec:juptel}. In Sect.
\ref{sec:jupvlbi}, correlations of simultaneous data recorded with
$\textsc{its}$ and $\textsc{nda}$ are described and then discussed
in Sect. \ref{sec:jup_concl&outlook}. Correlations were performed
on dynamic spectra and on the time domain data.

\section{Radio telescopes\label{sec:juptel}}

\subsection{ITS\label{subsec:jupits}}

$\textsc{lofar}$'s Initial Test Station was located at the core site
of $\textsc{lofar}$ near the village of Exloo, The Netherlands and
consists of 60 inverted-V-shaped dipole antennae distributed over
five spiral arms. $\textsc{its}$ antennae work in the lowest $\textsc{lofar}$
frequency band from 5 MHz to 35 MHz, which matches the emission band
of Jupiter-Io emission.

The final $\textsc{lofar}$ telescope will eventually consist of $\sim$15000
antennae, split up between a dense core of 32 stations and 45 remote
stations with 192 antennae each. These stations will be spaced on
exponentially expanding shells. The antennae within the station will
be placed on exponential shells with a maximum diameter of 60 meters.
The total bandwidth from 30 MHz to 240 MHz will be covered by two
different antenna types: a low band antenna ($\textsc{lba}$) for
frequencies below 80 MHz, and a high band antenna ($\textsc{hba}$)
from 110 MHz to 240 MHz. Below 30 MHz the gain of the low band antennae
is reduced, but a fixed hardware limit is at 10 MHz. The $\textsc{lofar}$
electronics is designed to accommodate a third type of antenna, which
could cover frequencies below 30 MHz to have a broader bandwidth for
Jupiter observations. With a maximum baseline of $\sim$100 km between
stations, all these antennae will provide $\textsc{lofar}$ with a
2 to 20 arcseconds angular resolution and millijansky sensitivity,
dependent on frequency. Each dipole's signal will be amplified in
an $\textsc{lna}$ (low-noise amplifier), filtered in frequency and
digitized \citep{falcke06}. More detail on $\textsc{lofar}$ and
its test stations can be found at \url{www.lofar.org}. Additionally,
a formulation of a science case with long baselines by extremely remote
LOFAR stations is described by \citet{vogt06}.

In the case of $\textsc{its}$, high performance analog-to-digital
converters ($\textsc{adc}s$) with a dynamic range of 12 bits, supply
a rate of 80 million samples per second. The sample rate results in
a time resolution of 12.5 ns and a data rate of $\sim$160 MB per
second and dipole. Four of those $\textsc{adc}s$ are mounted on two
$\textsc{tim}$-boards (twin-input module) which are accommodated
in each of the 16 standard $\textsc{pc}s$ in the data acquisition
($\textsc{daq}$) container. The $\textsc{tim}$-boards can handle
two digital data streams and store each stream in 1 GB of $\textsc{ram}$.
This amount of $\textsc{ram}$ limits the $\textsc{daq}$ to $6.7~\unit{s}$.
From 2005 on, 30 antennae measured dual polarization, since two orthogonal
dipoles were mounted in each antenna structure. Each antenna detects
the whole sky with a primary beam of about 90$^{\circ}$ full width
half maximum ($\textsc{fwhm}$) centered on the zenith. Using digital
beam-forming the array can be simultaneously pointed at several sources
in the sky. For $\textsc{its}$ the digital beam-forming was performed
off-line and is described in Sect. \ref{beamforming}. The absolute
timing of $\textsc{its}$ was updated with Network Time Protocol ($\textsc{ntp}$).
The beam size of $\textsc{its}$ at the zenith results in an angular
resolution of 3°, at a wavelength of 10 meters (30 MHz) and for a
maximum baseline of 180 meters.

\subsection{NDA\label{subsec:jupnda}}

The Nan\c{c}ay Decametric Array (\url{www.obs-nancay.fr/dam}) is
a phased array made up of 144 antennae. Each antenna consists of eight
conducting wires (monopoles) wound up on vertical support cables on
a conical surface. Each opposite pair of monopoles is connected as
a spiral dipole. The array is split in two groups of 72 left-handed
and 72 right-handed wound antennae to measure circularly polarized
radio waves. All 72 antennae of one cluster measure the same polarization
direction and can be phased up with an analog beam-former. A full
description of the array can be found in \citet{boischot80}.

The receiver used for the observations in 2005 provided the same type
of data as $\textsc{its}$ by sampling in the first Nyquist zone with
80 MHz over a frequency band from 8 MHz to 40 MHz. A $\textsc{daq}$
board (Signatec PDA14) was used for the digitization with a 14-bit
$\textsc{adc}$ inserted in a powerful $\textsc{pc}$, which allows
continuous $\textsc{daq}$ of a stream of 80 million samples per second.
These data are written as short integers (16-bit) to disc. The receiver
allows full signal detection below 40 MHz (a low-pass filter at 40
MHz and high-pass filter at 8 MHz, 16 MHz or 24 MHz are used) without
use of a local oscillator for frequency down conversion. Absolute
timing of the instrument is provided by a receiver of the Global Positioning
System ($\textsc{gps}$).

\section{Observations\label{sec:jupobs}}

Triggered by $\textsc{nda}$, we took simultaneous datasets during
periods of Jupiter burst emission. Simultaneous Jupiter snapshots
of $6.7~\unit{s}$ (1 GB) at $\textsc{its}$ and $20~\unit{s}$ ($\sim$3
GB) at $\textsc{nda}$ were recorded. The trigger was manually executed
at $\textsc{nda}$ while monitoring the real-time display of $\textsc{nda}$
for actual burst activity. The real-time monitor produces and displays
dynamic spectra at a rate of one spectrum per second. The monitor
can also be found online with a 30 second refreshment time (\url{www.obs-nancay.fr/dam/dam_visu.html}).
The day of observation was selected from $\textsc{nda}$'s monthly
probability plots (\url{www.obs-nancay.fr/dam/dam_proba.htm}, \citealt{leblanc93}).

Both telescopes recorded the full signal during overlapping periods.
The data taken with $\textsc{nda}$ are available in digital format
and contain the digitized output of the analog-formed beam in the
direction of Jupiter. For $\textsc{its}$ the raw digitized data for
all antennae were recorded. Total intensity was not calculated due
to uncertain relative gain calibration. Since a single dipole antenna
measured the whole sky at the same time, it detected also the sky
background as well as strong radio frequency interference ($\textsc{rfi}$).
The resulting high noise level makes it impossible to observe Jupiter's
emission in one antenna with enough contrast. The phasing of the antennae
by digital beam-forming, in the direction of Jupiter, was performed
to increase the signal-to-noise ratio by suppressing the $\textsc{rfi}$
coming from the horizon. 

To beam-form and display the data, the following steps were performed.
\label{beamforming}First, a Fourier transform, however the computation
time of the fast Fourier transform ($\textsc{fft}$) increases as
$N~logN$, where $N$ is the number of samples. Thus, splitting the
antenna data in time-blocks is more efficient than processing one
long time-series. However, the splitting requires extra bookkeeping
when reconstructing the beam-formed time-series needed for displaying
and for signal correlation.

Second, a Hamming window was applied to the raw data blocks to keep
the leakage in frequency domain into neighbouring channels as low
as possible. Third, the $\textsc{fft}$ was applied on blocks of 16384
samples. Fourth, the interesting band or bands were selected from
the frequency domain. Fifth, the beam-forming phases were applied
to each frequency channel. The beam-forming phases correct the differential
light travel time to each dipole-element relative to a reference position
or antenna. They are calculated for the exact fractional sample delay,
that shifts the antenna data for the exact desired beam direction.
The chosen block size determined a time resolution of 0.2 ms and a
spectral resolution of 5 kHz. After averaging all the phase-shifted
antenna spectra the resulting beam-spectra were stacked for each block
and displayed in a dynamic spectrum. The process of true time-delay
beam steering by phase-shifting and element averaging on $\textsc{adc}$
data is called digital beam-forming.

For the reconstruction of the time-series, the beam-formed blocks
had to partly overlap. The start and end of each beam-formed block
will not contain data from all antennae, due to the delays applied
to each antenna during beam-forming. For the resulting beam-formed
time-series, ultimately only intervals over which delay-shifted data
from all antennae exist are usable.

\begin{figure}
\begin{centering}\includegraphics[width=1\columnwidth]{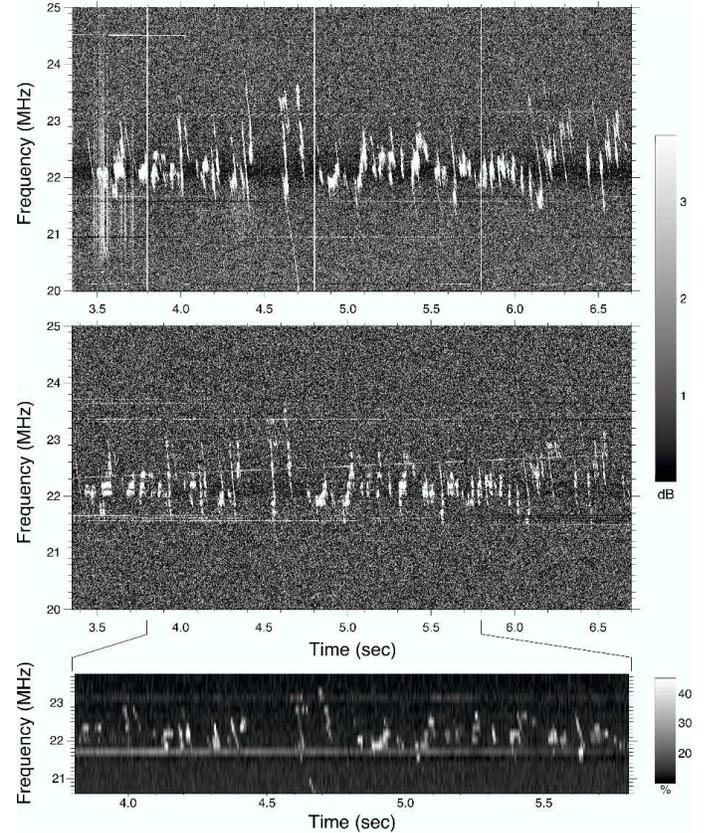}\end{centering}

\caption{\label{fig:jupX3_ds}Dynamic spectra of a simultaneous observation
with the Nan\c{c}ay Decametric Array ($\textsc{nda}$) {[}$\textsc{top}$]
and $\textsc{lofar}$'s Initial Test Station ($\textsc{lofar/its}$)
{[}$\textsc{center}$], on November 30, 2005 at 07:15:12 $\textsc{utc}$
for a duration of $\sim$3.5 s. Jovian S-bursts show up as the discrete
structures mainly confined to the band 20.7 MHz to 23.9 MHz. Dark
horizontal Faraday fringes are superimposed on emission observed at
$\textsc{its}$. Vertical lines in the $\textsc{nda}$ spectrum are
5 ms pulses intentionally added every second to the data for timing
purposes. Horizontal lines are caused by radio frequency interference
$(\textsc{rfi})$. The third panel {[}$\textsc{bottom}$] shows the
resulting cross correlation coefficients versus time and frequency,
computed from data blocks of 2.5 ms duration and $\sim$150 kHz bandwidth.
}
\end{figure}

\begin{figure}
\begin{centering}\includegraphics[width=1\columnwidth]{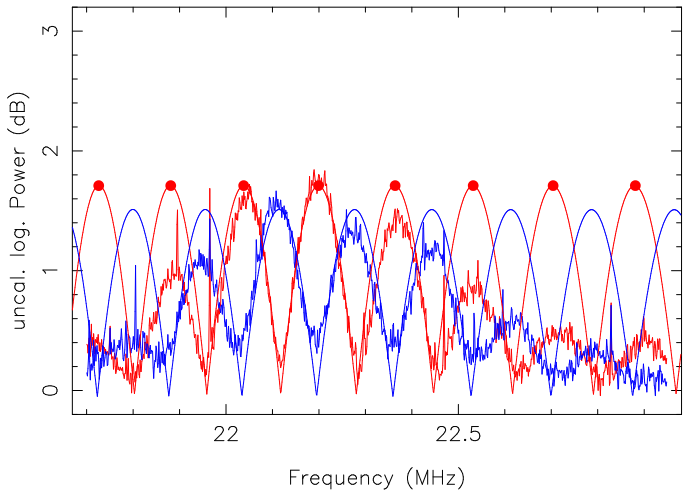}\end{centering}

\caption{\label{fig:jupX3_frot} Spectra of the two linear polarizations of
$\textsc{lofar}$'s Initial Test Station ($\textsc{lofar/its}$) {[}red:
E-W antennae dipole beam and blue: N-S antennae dipole beam]. The
best-fit sine function (in frequency, not intensity) of the computed
Faraday angle is superimposed to both $\textsc{its}$ spectra. For
the East-West polarization direction the maxima of the Faraday fringes
are indicated with dots. }
\end{figure}

For visual comparison of the simultaneous data, the dynamic power
spectra from the two telescopes are plotted in Fig. \ref{fig:jupX3_ds}.
The 48 antennae of $\textsc{its}$ and the 72 antennae of $\textsc{nda}$
were beam-formed in the direction of Jupiter. The spectra were computed
with 0.2 ms time resolution and consequently a spectral resolution
of 4.9 kHz. The Jupiter signal received was expected between 20 MHz
and 30 MHz, where only a few narrow band lines contaminate the spectra.
Below 20 MHz strong $\textsc{rfi}$ is dominating the spectra and
above 30 MHz the bandpass filter of $\textsc{its}$ rolls off. The
Jupiter-Io S-burst emission is clearly visible between 20.7 MHz to
23.9 MHz and several identical features can be identified. The emission
drifts in frequency, which translates to a movement of the source
region along Jupiter's magnetic field lines. These S-bursts last for
a fraction of a second down to a few milliseconds. The horizontal
bands in the $\textsc{its}$ plot are due to Faraday rotation. The
power spectra of the two perpendicular polarization directions of
$\textsc{its}$ show strong Faraday fringes (see Fig. \ref{fig:jupX3_frot}).
The spectra have a spectral resolution of 1.2 kHz and are integrated
over $6.7~\unit{s}$. The plotted frequency range from 21.7 MHz to
22.95 MHz covers the strong Jupiter burst emission and contains only
little $\textsc{rfi}$ (narrow vertical lines). The Faraday fringes
were fitted with a sinusoidal function according to Faraday's law.
The resulting rotation measure of $RM=(1.15\pm0.07)~\unit{rad~m^{-2}}$
is in good agreement with the rotation measure determined from total
electron content values provided by the $\textsc{gps}$. For the East-West
polarization direction, the maxima of the Faraday fringes are indicated
with dots. The width of the Faraday fringes was measured as 100 kHz
to 150 kHz. Around the frequencies of the eight maxima, the signal
was filtered for the cross correlations described in Sect. \ref{subsec:jupxcor}.
For a more detailed analysis of the Faraday rotation and previous
results see \citet{nigl05previ}.

\section{Long baseline interferometry\label{sec:jupvlbi}}

For this work, Jupiter was observed with $\textsc{its}$ and $\textsc{nda}$
on a baseline $D$ of $702~\unit{km}$, corresponding to an angular
resolution $\theta$ of $3.3^{"}$ at a frequency of $\nu=30~\unit{MHz}$.
For this resolution, Jupiter burst emission regions can be considered
as point sources. The smallest structure that could be resolved on
Jupiter at the time of the observation was $13600~\unit{km}$ in diameter,
which is four times the size of Io.

\subsection{2D Cross correlation of dynamic spectra\label{subsec:jup2dsxcor}}

The synchronization of the absolute timing of $\textsc{its}$ and
$\textsc{nda}$ data was of the order of one second. In order to restrict
the size of the time window for correlation, a normalized 2D cross
correlation of the dynamic spectra shown in Fig. \ref{fig:jupX3_ds}
was performed. This cross correlation provided a coarse fringe search
for the time delay and an offset in frequency between the two datasets,
which is called a clock model in $\textsc{vlbi}$ terminology. Before
correlation, the dynamic spectra have to be generated with high time
resolution for measuring the time delay, and with high spectral resolution
for determining the frequency offset. The higher the resolution in
time, the lower in frequency, maintaining a time-bandwidth product
of unity ($d\nu\times dt=1$). Thus for high time resolution, fine
spectral structure is lost, which reduces the quality of the subsequent
2D cross correlation, but gives access to an accurate time delay.
An offset in frequency can be found by choosing high spectral resolution.

\begin{figure}
\begin{centering}\includegraphics[width=1\columnwidth]{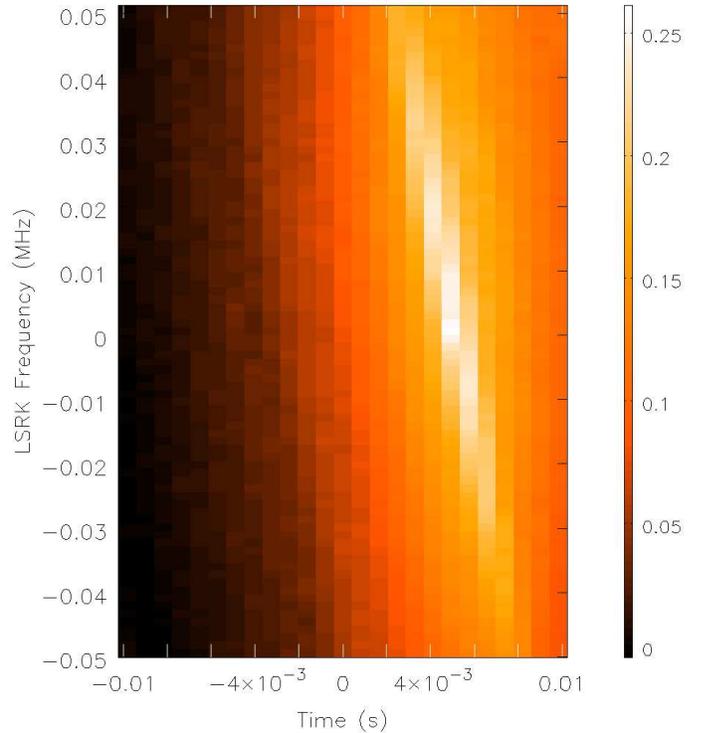}\end{centering}

\caption{\label{fig:dsxcores} Correlation coefficient matrix for two dynamic
spectra slices. The intensity in the image displays the cross correlation
coefficient labeled with the color wedge on the right side of the
plot. }
\end{figure}

For the observation performed on November 30, 2005 at 07:15:12 $\textsc{utc}$,
the result of a 2D cross correlation, of two slices of the dynamic
spectra from $\textsc{its}$ and $\textsc{nda}$ is given in Fig.
\ref{fig:dsxcores}. In this correlation window, maximum correlation
is found at zero frequency shift and at a time delay of $4.9~\unit{ms}$.
The clock offset resulted in $10.0723328~\unit{s}$ with a resolution
of $1.6~\unit{\mu s}$ ($\textsc{its}$ started to record later than
$\textsc{nda}$). This clock offset (clock model) serves as input
for the time-series cross correlation. Additionally, a shift in frequency
of $191~\unit{Hz}$ with a resolution of $38~\unit{Hz}$ was obtained,
which will be explained in Sect. \ref{sub:jup_xcor_lags} as being
mainly due to a difference between the sample rate of both telescopes,
caused by the deviation of the $\textsc{its}$ system from the nominal
sample rate.

The sample rate was measured for $\textsc{nda}$ with $(79999998\pm0.5)$
samples per second differing only slightly from the nominal sample
rate of $80$ megasamples per second. This difference causes an offset
from the expected and reconstructed frequency in the dynamic spectrum
of $+0.6~\unit{Hz}$ (for $\textsc{nda}$ at $22.3~\unit{MHz}$).
Furthermore, due to Earth rotation, the radial speed of the two telescopes
with respect to Jupiter at the time of the observation differed by
$38~\unit{m~s^{-1}}$, with $\textsc{its}$ moving towards Jupiter
as seen from $\textsc{nda}$. Therefore the $\textsc{its}$ signal
was stronger Doppler shifted to higher frequencies by $2.8~\unit{Hz}$.
Both frequency offsets lie within the frequency offset resulting from
the 2D cross correlation above.

\subsection{Coherence time and coherence bandwidth\label{subsec:juplimits}}

Since emission from Jupiter arrives in bursts, the best signal-to-nose
ratio is achieved by correlating with integration time and frequency
resolution matched to the duration and bandwidth of the bursts. The
emitting electron bunches at Jupiter are traveling along the $\textsc{ift}$
at several thousands to several tens of thousands of kilometers per
second. The emission is received as subpulses (wave packets) with
about $40~\unit{\mu s}$ to $90~\unit{\mu s}$ duration and a minimum
bandwidth of $30~\unit{kHz}$, as it has been observed by \citet{carr99}.
Therefore the blocks for correlation should be much longer than a
wave packet of $90~\unit{\mu s}$ (thus $\simeq1~\unit{ms}$), and
have a frequency resolution coarser than $30~\unit{kHz}$, but not
more than the Faraday fringe spacing of $150~\unit{kHz}$. The upper
limit for the correlation time window is dependent on the time-scales
of changes in the Earth ionosphere and has yet to be identified. The
discussed limits constrain the optimum window for the cross correlation.

\subsection{Cross correlation of time series\label{subsec:jupxcor}}

The final cross correlation was done in the frequency domain following
Parseval's Theorem. The fringe amplitude was calculated with the so-called
fringe correlation method as the square root of the sum of the squares
of the quadrature correlations \citep[compare][]{lynch76}. The cosine
component was obtained by a second cross correlation with a quarter-period
shift at the center frequency of the selected subband introduced to
one of the time series. The quarter-period shift was applied in the
frequency domain as a phase-gradient, which enables shifts by a fraction
of a time sample.

We used the cross correlation theorem, which performs the cross correlation
of one block, for all possible lags, by only three Fourier transformations
and one multiplication of the two data arrays. The lags in the cross
correlation result in sample steps of $12.5~\unit{ns}$.

\begin{figure}
\begin{centering}\includegraphics[width=1\columnwidth]{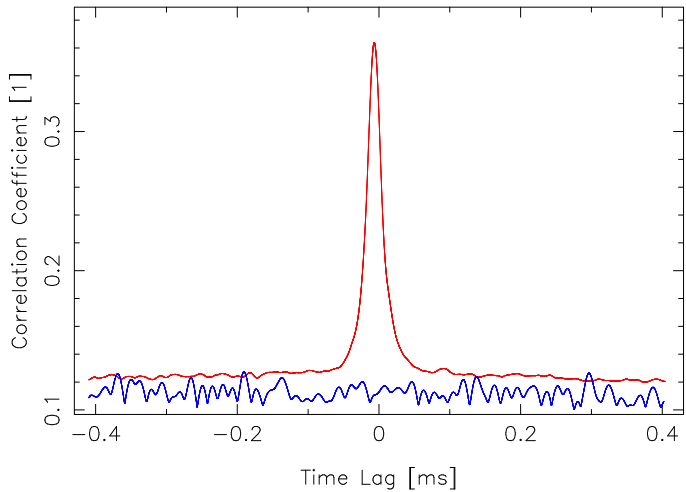}\end{centering}

\caption{\label{fig:jup_wfcor}Result of the time-series signal cross correlation
of the $\textsc{nda}$ and $\textsc{its}$ data sets of the Jupiter
observation on 30.11.2005 at 07:15:12 $\textsc{utc}$. The (red) peaked
curve in the plot was obtained by averaging about 1000 correlated
1 ms time-series blocks containing Jupiter burst signal at eight bands
of $75~\unit{kHz}$ bandwidth. The (blue) flat curve was obtained
on the same number of blocks without emission. The peak maximum lies
210 sigma above the noise level with 1 s effective integration. }
\end{figure}

Since the observed Jupiter-Io emission is not continuous, noise gating
was applied and only time blocks with received power exceeding the
noise-floor were added after cross correlation according to the fringe
correlation method. Integrating about one thousand 1 ms blocks containing
burst emission at eight bands of $75~\unit{kHz}$ bandwidth, resulted
in an average cross correlation peak of 210 sigma significance compared
to the same number of integrated blocks containing only background
noise (see Fig. \ref{fig:jup_wfcor}).

In total, narrow bands with bandwidths of $(30$, $75$ and $150)~\unit{kHz}$
at center frequencies of $(21.73$, $21.88$, $22.04$, $22.20$,
$22.36$, $22.53$, $22.70$ and $22.88)~\unit{MHz}$ were analyzed.
These frequencies were chosen as they lie at the strong maxima of
the Faraday fringes from the $\textsc{its}$ power spectra, which
are shown in Fig. \ref{fig:jupX3_frot}.

The bottom panel in Fig. \ref{fig:jupX3_ds} displays the cross-correlation
coefficients in a dynamic spectrum for 2 seconds between 20.6 and
23.8 MHz, computed from 2.5 milliseconds time blocks over all bright
Faraday fringes.

The distribution of cross correlation coefficients of 500 correlated
1 millisecond time blocks (selected by noise gating), was maximum
for a bandwidth of $30~\unit{kHz}$ at a center frequency of $22.53~\unit{MHz}$
resulting in $C=0.6\pm0.1$ (one sigma statistical error stated).

\subsection{Fringe visibilities}

The cross correlation coefficients of the time blocks are proportional
to the fringe visibilities, which ultimately carry information on
the source size, following for example \citet{phillips87}:\begin{eqnarray}
\gamma_{S} & = & \frac{1}{G_{\tau}}kC\eta\sqrt{\frac{(P_{S1}+P_{B1})(P_{S2}+P_{B2})}{P_{S1}P_{S2}}}\label{eq:visibility}\end{eqnarray}

Here $\gamma_{S}$ is the instrument-corrected fringe visibility due
to the source structure; $G_{\tau}$ is the fringe washing factor
at the residual geometrical delay $\tau$; $C$ is the cross correlation
coefficient; $k$ and $\eta$ are instrumental decorrelation coefficients;
$P_{S}$ is the power from the source and $P_{B}$ the power from
the Galactic background both recorded at the two telescopes. The instrumental
decorrelation coefficient $\eta$ accounts for any phase behavior
of filters, amplifiers, cables and connectors of both telescopes together.
For $\textsc{its/nda}$ this factor was assumed to be maximum 15\%
and thus $\eta=1.15$.

Since only circular polarized parts of the Jupiter signal were measured
by $\textsc{nda}$ and correlated with only linear polarized components
of $\textsc{its}$, a maximum cross correlation coefficient of $C_{max}=\sqrt{2}/2$
can be achieved. Therefore the second instrumental coefficient was
adopted to $k=\sqrt{2}$.

The fringe washing function, $G_{\tau}$, is the Fourier transform
of the receiver passband \citep{thompson86}. For a rectangular passband,
$G_{\tau}$ is a sinc function: $G_{\tau}(\tau)=sinc(\pi\Delta\nu\tau)$.
Here, $\Delta\nu$ is the bandwidth of the correlated signal and $\tau$
is a residual geometrical delay, which is the difference between the
true delay and the actually applied delay for the correlation: $\tau=\tau_{g}-\tau_{i}$.
In the off-line correlations performed for this paper, the lag of
the maximum correlation was determined for each block, which resulted
in $\tau=0$ and $G=1$. The two terms $(signal+noise)/signal$ under
the square root in Eq. $\ref{eq:visibility}$ are estimated to be
$\sim1.04$ for $\textsc{nda}$ and $\sim1.06$ for $\textsc{its}$.
A global background correction factor of $1.1$ is obtained. The resulting
visibilities $\gamma_{S}$ are consistent with one, and thus with
unresolved S-burst sources, as expected.

\subsection{Cross correlation lags\label{sub:jup_xcor_lags}}

The lag in which the correlation maxima laid, and hence the signal
arrival time difference between the two datasets, changed by $(-8.8\pm0.1)~\unit{\mu s~s^{-1}}$
(one sigma statistical error stated). The minus sign means that the
signal arrived progressively earlier at $\textsc{its}$. This drift
in the lag is mainly caused by a constant relative drift of the clock
quartz of $\textsc{its}$ with an accuracy of $\Delta F/F=10^{-4}$.
Here $F$ is the nominal frequency and $\Delta F$ the maximum deviation
including errors due to calibration, temperature variation, shock,
vibration and aging over 10 years. The accuracy of the $\textsc{nda}$
clock is much better and was measured by $\textsc{gps}$ with $\Delta F/F=1.25\times10^{-8}$
and thus accounts for a maximum drift rate of only $12.5~\unit{ns~s^{-1}}$.
The motion of the source with an average parallel electron velocity
of $\sim22000~\unit{km~s^{-1}}$ \citep{zarka96} translates to a
drift of only $-367.82~\unit{ns~s^{-1}}$ at the Earth. Interplanetary
multipath scintillation can cause delays of the order of only a few
nanoseconds assuming a thin scattering sheet \citep{phillips87}.
The Earth rotation during the short observation only accounts for
$+125.17~\unit{ns~s^{-1}}$ and ionospheric effects change on time-scales
longer than the duration of our observation and therefore were approximated
by a differential change in the local slant $\textsc{tec}$ values,
resulting in a contribution of $-0.032~\unit{ns~s^{-1}}$.

Additionally, the drift of the $\textsc{its}$ clock causes a deviation
between the sample rates of $\textsc{nda}$ and $\textsc{its}$ of
$704~\unit{Hz}$. The bigger the time-window for the cross correlation,
the larger the deviation in time per correlated sample gets, which
reduces the quality of the correlation. Furthermore, this difference
in the sample rates translates into an offset in frequency of $196~\unit{Hz}$
between the dynamic spectra of both telescopes, which is consistent
with the frequency offset of $191~\unit{Hz}$ obtained from the 2D
cross correlation of the dynamic spectra with a resolution of $38~\unit{Hz}$
(Sect. \ref{subsec:jup2dsxcor}).

\section{Conclusions \& outlook\label{sec:jup_concl&outlook}}

Using the prototype $\textsc{its}$ of the new generation radio telescope
$\textsc{lofar}$ and the Nan\c{c}ay Decametric Array, strong Io-induced
Jupiter radio emission, including S-bursts, was detected over 6.7
seconds with only dozens of antennae. The beamed dynamic spectra obtained
with $\textsc{its}$ and $\textsc{nda}$ show excellent agreement
(see Fig. \ref{fig:jupX3_ds}). High time and frequency resolution
allowed us to determine offsets by two-dimensional cross correlation
of the dynamic spectra with a precision of microsecond accuracy in
time and hertz accuracy in frequency. The latter method revealed a
shift in frequency, due to the relative radial speed and an offset
in sample frequency between the two telescopes.

The baseline of the interferometer of 702 km provided an angular resolution
of three arcseconds. Since the emitting source region is believed
to be smaller than 0.02 arcseconds \citep{zarka96}, the source region
of the received emission remains unresolved, as confirmed by the obtained
visibilities being close to one. The identified delays at the maxima
of the time-series cross correlations showed a linear drift, caused
by a large clock rate at $\textsc{its}$. Correcting for this constant
drift, no significant variation was found within the observation of
$6.7~\unit{s}$. One second effective integration in time and several
hundred kilohertz bands in frequency resulted in an 210 sigma detection
above the noise level.

This is the first time that $\textsc{vlbi}$-type observations have
been performed over a broad relative bandwidth with baseband digitization,
and direct correlation of signals with a large dynamic range. Cross
correlation fringes on the beam-formed time-series were found on time-scales
of milliseconds. As demonstrated with the $\textsc{its}$ prototype,
$\textsc{lofar}$ will probably reach arcsecond angular resolution.
$\textsc{lofar}$ will be pioneering in being capable of performing
$\textsc{vlbi}$-type observations as described in this article with
millijansky sensitivity. The present results support the European
extension of $\textsc{lofar}$ with very remote stations up to several
hundred kilometers. A baseline of a thousand kilometers will serve
$\textsc{lofar}$ with a spatial resolution of an arcsecond at 30
MHz. Such a resolution will enable detailed studies of the Jupiter
magnetosphere as described in \citet{zarka04} among many other astrophysical
applications \citep{vogt06}.

A follow up of these long-baseline interferometry studies is planned
with $\textsc{nda}$ and $\textsc{lofar/cs1}$ (Core Station 1). The
follow up experiments will be performed to quantify at what duty cycle
time-phase-coherence is preserved, as a function of time and frequency.
Furthermore, weaker radio sources may be used in addition to Jupiter.

\begin{acknowledgements}
We wish to thank for their contribution to the observations and results presented here, the team at \textsc{astron} for operating \textsc{lofar/its}, no matter what day, time or obstacle and Sebastien Hess at the Observatoire de Paris, who helped us to perform the observations with \textsc{nda}.
\end{acknowledgements}

\bibliographystyle{aa}
\bibliography{anigl}

\end{document}